\documentclass[12pt]{iopart}

\usepackage{iopams}
\usepackage{graphicx}
\usepackage{enumerate}
\graphicspath{{Figurer/}}

\renewcommand{\AM}{\textbf{A}}

\newcommand{\FM}{\textbf{F}}
\newcommand{\rM}{\textbf{r}}
\newcommand{\vM}{\textbf{v}}
\newcommand{\aM}{\textbf{a}}
\newcommand{\eM}{\textbf{e}}
\newcommand{\BM}{\textbf{B}}
\newcommand{\al}{\alpha}
\newcommand{\en}{\epsilon}
\newcommand{\ta}{\theta}
\newcommand{\ti}{\theta_i}
\newcommand{\te}{\theta_e}

\newcommand{\hm}{\widehat m}       
\newcommand{\D}{\Delta}
\newcommand{\F}{\mathcal{F}}
\newcommand{\I}{\mathcal{I}}
\newcommand{\M}{\mathcal{M}}
\newcommand{\tL}{\widetilde{L}}
\newcommand{\tm}{\widetilde{m}}
\newcommand{\mM}{\widetilde{\textbf{m}}}

\begin{document}

\title[Experimental results on inner motion]
{The swinging counterweight trebuchet\\
Experiments on inner movement and ranges}

\author{Erik Horsdal, Filip Drejer Johnsen and Jonas Rasmussen}
\address{Department of Physics and Astronomy, Aarhus University,
DK-8000 Aarhus C, Denmark}
\ead{horsdal@phys.au.dk}
\begin{abstract}	
	The inner movement of a trebuchet with swinging counterweight was measured 
	inside a laboratory	by the use of rotation sensors to determine angular coordinates 
	for beam, counterweight and sling.
	Data collection was started before the trebuchet was fired and lasted
	until some time after the projectile was released
	and flung into ground right in front of the machine.
	A single measurement is then sufficient for a semi-empirical determination of
	longest range and kinetic energy at target as well as mechanical energies and 
	internal forces throughout the entire shot. 
	Loss of mechanical energy was ascertained and attributed to friction.
	Measurements of actual ranges performed on a flat field with little wind are presented 
	and compared with the semi-empirical determinations and ab initio calculations.
\end{abstract}

%
\section{Introduction}
Trebuchets are historical engines of war.
At their peak, they hurled heavy stones at castles under siege from long 
distances and caused great destruction.
Discussions of the use, development, and physics of these artillery pieces can 
be found in~\cite{ref:PVH,ref:Chev,ref:TS,ref:Chev_et_al,ref:MSF1,ref:MSF2}. 
There is little doubt that the trebuchet became the most important siege weapon 
during a period of the late Middle Ages before it was overtaken by cannons and 
gunpowder, but in spite of this, only little information exists on the precise 
achievements and dimensions of historical trebuchets, and a generally accepted 
standard for their performance is also lacking~\cite{ref:Horsdal}.
Aside from their continuing historical interest, trebuchets also have
enjoyed some attention in practical teaching of classical
mechanics~\cite{ref:O'Connell,ref:Denny_2005,ref:Denny_2009,ref:Christo}
because the internal movement touches on many important concepts.

A relatively small experimental trebuchet that could be handled safely by one 
or two persons inside a laboratory was built and equipped with sensors
for the measurement of all angular movements during a shot.
The position~$\rM$ and velocity~$\vM$ of the projectile at any time prior to 
its release can be calculated from these data and perceived as initial 
conditions for a ballistic trajectory from which the range can be extracted
as a function of release time or position.
We focus on longest range and the energy at target that follows, 
but internal forces, mechanical energies and their reduction due to internal 
friction can also be calculated both before and after the projectile is released.

The trebuchet was taken out in the open on warm and calm days for 
the measurement of actual ranges on a flat field.
Two sling lengths and two projectiles in the form of natural, round stones 
were used.
Many shots had to be fired before a maximum range and its random error could be 
determined.
Some were carried out with different settings of the release mechanism, 
and others as repetitions under identical conditions, as far as possible,
to expose random errors. 
\section{Experimental arrangement and procedures}
\subsection{The trebuchet and angular measurements}
\begin{figure}[htb]
	\centering	
	\includegraphics[width=0.80\textwidth]{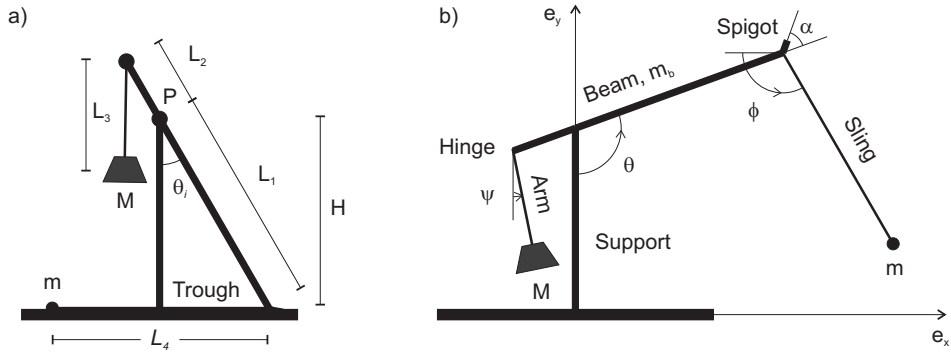}
		\caption{
		a)~Initial configuration, projectile in trough.
		b)~Shortly after lift-off.}
		\label{fig:Trebuchet}
\end{figure}

Schematic diagrams of the trebuchet are shown in~\fref{fig:Trebuchet}.
The linear dimensions identified in~\fref{fig:Trebuchet}a are
the long and short segments of the beam~$L_1$ and~$L_2$, respectively,
the length~$L_3$ of the arm that holds the counterweight, and the sling 
length~$L_4$.
The height of the pivoting axle~$P$ for the beam is~$H$, and~$m$ and~$M$ 
are the projectile and counterweight masses, respectively.
Rotation sensors on the engine measure the respective angular 
coordinates~$\ta$,~$\psi$ and~$\phi$ of beam, counterweight, and sling 
as functions of time.
These angles and their orientations are shown in~\fref{fig:Trebuchet}b.
Two are measured from the vertical direction~$(\ta,\psi)$ and one from the
horizontal~$\phi$.

\begin{figure}[htb]
	\centering	
	\includegraphics[width=0.75\textwidth]{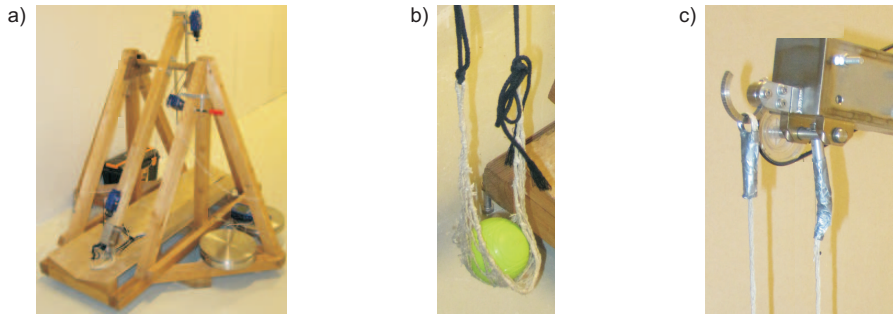}
		\caption{a)~Trebuchet with beam in initial position.
								Three blue rotation sensors.
						 b)~Projectile and pouch.
						 c)~Two cords of sling. One hanging from semicircle.}
		\label{fig:TrebuchetPhoto}
\end{figure}

\begin{figure}[htb]
	\centering	
	\includegraphics[width=0.65\textwidth]{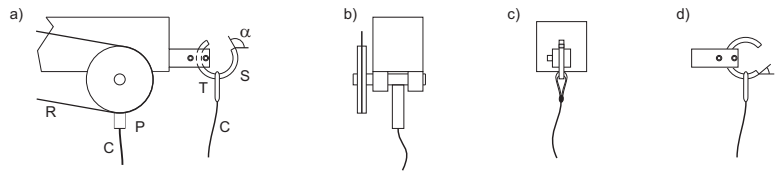}
		\caption{
		Beam and attachment of sling.
		a)~Side view.
		Cord~C connected to pin~P.
		Second cord~C with ring~T, semicircular spigot~S and angle~$\al$.
		Rubber band~R connects to sensor on beam.
		b)~End view without spigot. Pin, axle, two ball bearings, and wheel.
		c)~End view showing spigot.
		d)~Small spigot angle.}
		\label{fig:SlyngeSpigot}
\end{figure}

Photographic images are found in \fref{fig:TrebuchetPhoto}.
The configuration before a shot is displayed in~\fref{fig:TrebuchetPhoto}a, 
but the sling is empty and the counterweight is not mounted.
It consists of a few stainless steel discs, each weighing~$\simeq$10kg, 
and two are seen in the foreground on the base next to the trough and trestle.
Three blue rotation sensors are also visible.
\Fref{fig:TrebuchetPhoto}b shows the pouch of the sling with a projectile 
(tennis ball), and in \fref{fig:TrebuchetPhoto}c one sees
the spigot (open end of adjustable semicircle) and the two cords of the sling.
One is tied permanently to a pin welded perpendicular onto a stainless steel 
axle that carries a wheel at the far end for angular measurement.
The other is tied to a ring held by the semicircle on which it slides as a shot 
progresses, and as soon as the direction of the sling becomes perpendicular to the 
spigot, the ring slides off and the pouch opens.
This is the release mechanism for the projectile.
Details are given in~\fref{fig:SlyngeSpigot}a and~\ref{fig:SlyngeSpigot}b.
The axle is supported by low friction ball bearings, and the wheel is coupled by 
a rubber band to an identical wheel on a rotation sensor sitting on the beam. 
The spigot may be fixed in different positions for different spigot angles~$\al$.
A large is seen in~\fref{fig:SlyngeSpigot}a and a small 
in~\fref{fig:SlyngeSpigot}d.
The length of the long beam section~$L_1=97.5$cm is the distance from pivot to 
midway between axle and ring as in~\fref{fig:SlyngeSpigot}a,
and the sling length~$L_4=87.0$cm is the vertical distance from the lower side 
of the beam in horizontal position to a heavy projectile that stretches 
the pouch.
The distance from axle to ring varies during a shot but is always close to~7cm. 
The projectile therefore moves relative to the beam almost on an ellipse 
with the focal points a distance~$d=7$cm apart. 
The eccentricity~$e$ and semi-major axis~$a\simeq2L_4$ are related to~$d$ 
by~$d=2ea$, so~$e\simeq0.04$.
This ellipse is almost indistinguishable from a circle of radius~$L_4$ and 
center at the midway point.
This approximation is used hereafter.

\begin{figure}[htb]
	\centering	
	\includegraphics[width=0.90\textwidth]{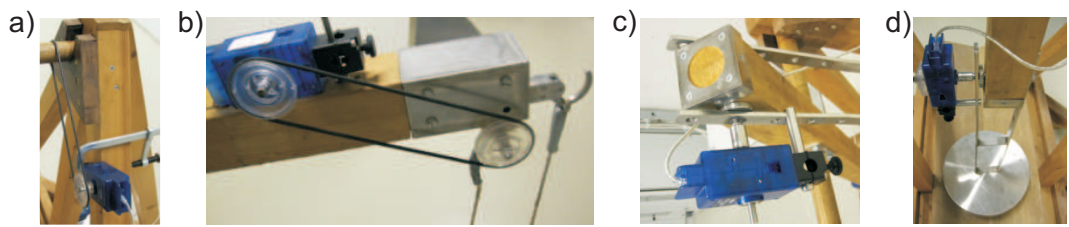}
		\caption{
		Rotation sensors for:
		a)~Beam~$\ta$.
		b)~Sling~$\phi$.
		c-d)~Counterweight~$\psi$.}
	\label{fig:SensorPhotos}
\end{figure}

The positions of the three rotation sensors and their couplings are shown in
\fref{fig:SensorPhotos}.
All sensors are designed to start an angular measurement from zero and to
distinguish between left and right rotations by including a sign which is adjusted
subsequently to comply with the conventions in~\fref{fig:Trebuchet}.
The first sensor, seen in~\fref{fig:SensorPhotos}a, is clamped to the trestle
and coupled to the pivoting axle of the beam by a rubber band.
The experimental beam angle~$\te$ is thus related to the coordinate angle~$\ta$ by
\begin{eqnarray}\label{eq:theta_e}
	\ta=k\te+\ti,
\end{eqnarray}
where~$k$ is a calibration constant that depends on the diameters of wheel and 
axle, and~$\ti$ is the initial value of~$\ta$.
The constant~$k$ was determined by varying the orientation of the beam from 
horizontal to vertical and measuring the resulting variation of~$\te$.
The correct variation of~$\ta$ by~$\pi/2$ was ensured by the use of 
sensitive spirit levels.
The initial beam angle~$\ti=0.55$rad was measured by 
lowering the beam to its initial position from a horizontal orientation
ensured again by a sensitive spirit level.

Another sensor, seen fixed to the beam in \fref{fig:SensorPhotos}b and coupled 
to the sling as explained above, monitors the angle between beam and 
sling.
The experimental angle~$\phi_e$ measured by the sensor is related to the coordinate
angle~$\phi$ for the sling by
\begin{eqnarray}\nonumber
	\phi=\phi_e+k\te.
\end{eqnarray}

Figures~\ref{fig:SensorPhotos}c-\ref{fig:SensorPhotos}d show the third sensor
at rest before and after a shot.
Its body is fixed to the arm that holds the counterweight and its rotation 
axle is clamped to a fixed axle that extends from the beam.
This sensor monitors the angle between beam and counterweight, and the
measured angle~$\psi_e$ is related to the coordinate angle~$\psi$ by
\begin{eqnarray}\label{eq:psi_e}
	\psi=\psi_e+k\te.
\end{eqnarray}
\subsection{Results of angular measurements}
\begin{figure}[htb]
	\centering	
	\includegraphics[width=0.60\textwidth]{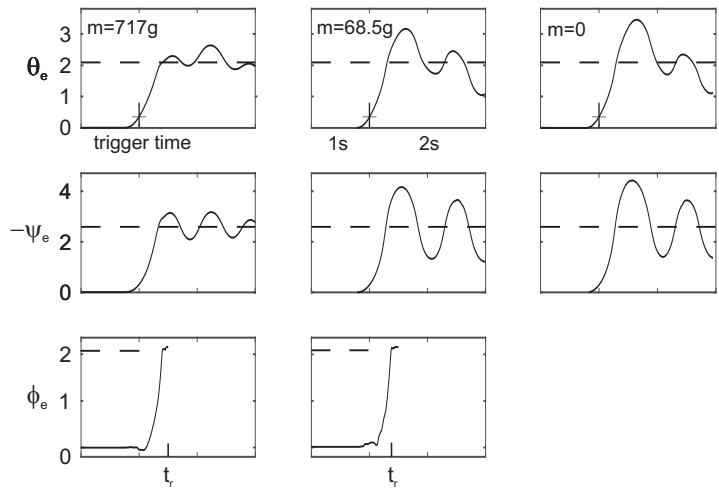}
		\caption{
			Angular measurements for three loads.
			Measurements triggered at crosses where~$\te=0.35$rad.
			Release time at~$t_r$ soon after pin stop at~$\phi_e=\pi/2+\ta_i=2.12$rad.}
		\label{fig:Prim_data}
\end{figure}

Angles measured as a function of time for beam, counterweight, and sling are 
shown in~\fref{fig:Prim_data} 
for three loads: 
A heavy metallic petanque ball to the left, a light tennis ball in the 
middle and a blank shot with no projectile to the right.
The sensors were set to sample at a rate of 100Hz and to use a pre-trigger 
buffer such that data are collected from one second before~$\te$ passes~$20^\circ$ 
and to stop two seconds later. 
The three seconds of acquisition time delivers~300 samples from well before the 
trebuchet is fired and until well after the projectile has left the pouch.
All angles start at zero and broken lines show final values at rest
for beam and counterweight.
Oscillatory motion is seen in each case. 
It is strongest in the blank shot to the right, a little damped by the 
light projectile, 
and quite significantly suppressed by the heavy, which carries a 
large fraction of the available mechanical energy after release.
The oscillations with smaller amplitude have larger frequency as expected 
for pendulum motion.

The two measured sling angles in the lowest row of~\fref{fig:Prim_data}
are meaningful only from some time after a shot is initiated and until 
the pin~P in~\fref{fig:SlyngeSpigot}a becomes parallel to the beam and 
can move no further. 
The beam was held at rest prior to each shot by a firm grip of the 
projectile by the operator.
The static tension in the cords of the sling therefore drops suddenly 
as soon as the projectile is set free, and the pin then falls a little to give 
a false indication of sling direction.
However, the tensions soon reach values well beyond the gravity~$mg$ of 
the projectile and this makes the measurements credible,
but the motion of the pin is stopped before the projectile is released. 
This is seen clearly in the data when the sharp increase of~$\phi_e$ is 
suddenly interrupted.
\subsection{Initial motion and moment of inertia for beam}
\begin{figure}[htb]
	\centering	
	\includegraphics[width=0.55\textwidth]{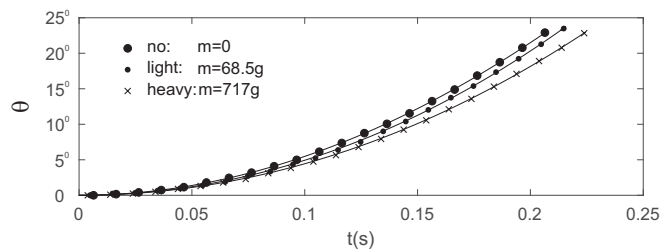}
		\caption{\label{fig:LH_t0}
		Initial variation of~$\ta$~\textit{vs} adjusted time.
		Full curves, fits determining~$\ddot\ta_i$.}
\end{figure}

The initial motion of the beam before triggering is shown in~\fref{fig:LH_t0} 
for the loads in~\fref{fig:Prim_data}.
The start of each shot at~$t=t_i$ was found by fitting the function
\begin{eqnarray}\label{eq:fitfu}
	f(t)=\frac{1}{2}\ddot\ta_i(t-t_i)^2+c(t-t_i)^3
\end{eqnarray}
to the measured angles after calibration and exclusion of the initial zeros.
Newtonean dot-notation is used here and henceforth for time differentiation.
The fitting parameters in~\eref{eq:fitfu} are, besides~$t_i$, the initial 
angular acceleration~$\ddot\ta_i$ and~$c$.

After replacement of~$t-t_i$ by~$t$, equation~\eref{eq:fitfu} reads
\begin{eqnarray}\nonumber
	\ta=\frac{1}{2}\ddot\ti t^2 + c t^3
	\quad\mathrm{and}\quad
	ct\ll\ddot\ti
	\quad\mathrm{at}\quad
	t=0.2s.
\end{eqnarray}
This is the function shown in~\fref{fig:LH_t0}.
It fits the data very well so the parameters~$t_i$ and~$\ddot\ti$ are 
determined with good precision.
The light projectile damps the acceleration a little relative to a blank shot, 
and the heavy projectile lowers it significantly.
We find~$\ddot\ti=19.2$rad/s$^2$ and~16.0rad/s$^2$ for the blank shot and the 
heavy projectile, respectively.
The initial acceleration of the projectile is~$L_1\cos\ti\ddot\ti$, which is
somewhat larger than the gravitational acceleration~$g$ in both cases.

Similar measurements and analysis with the beam alone, starting from a
horizontal position ensured by a sensitive spirit level,
gives~$\ddot\ta_{ih}=(-12\pm1)$rad/s$^2$, and this quantity is related to the moment 
of inertia~$\I_b$ for the beam relative to the pivoting axle by
\begin{eqnarray}
	\I_b\ddot\ta_{ih}=-m_bgL_{CM},
	\quad\mathrm{Newton's~second~law~for~rotation,}
	\nonumber
\end{eqnarray}
where~$m_b$ is the beam mass and~$L_{CM}$ the pivot to center of 
mass distance.
These two parameters were determined by weighing and balancing the 
beam with all its attachments.
This leads to~$\I_b=1.85\pm0.15$kgm$^2$.

\Tref{tab:Param} collects all parameters of the trebuchet,
\begin{table}[tbh]\footnotesize   
	\centering
		\begin{tabular}{ccccc|cccc|cc}
			$L_1$ & $L_2$ &	$L_3$ & $L_4$ & $M$  &
			$L_{CM}$	& $m_b$ & $\I_b$ &$\ti$	& $H$ & $\D U$			\\
			cm    & cm    & cm    & cm 		& kg 		&
			cm 		& kg    & kgm$^2$ & rad	& cm & J								\\\hline
			97.5	& 25.0	& 51.5					& 87.0	& 53.9 &
			46.5  & 4.86  & $1.85\pm0.15$ & 0.55 	& 83.1 & 204 
		\end{tabular}
	\caption{Parameters of trebuchet.
		$M$ includes~5 discs and their movable support.}
	\label{tab:Param}
\end{table}
and includes also the calculated height of the fulcrum~$H=L_1\cos\ti$ 
and the potential energy invested when the engine is loaded
\begin{eqnarray}\label{eq:DU}
	\D U=(ML_2-m_bL_{CM})g(1+\cos\ti).
\end{eqnarray}
Among these parameters, the moment of inertia~$\I_b$ is considered the most uncertain.
The semi-empirical ranges and energies at target extracted from the angular 
measurements do not depend on the value of~$\I_b$ in~\tref{tab:Param}, 
but it is important for the determination of dynamic variables like the mechanical 
energy of the engine and internal force.
The measured value of~$\I_b$ was therefore investigated further, see~\ref{app:Model}.
\subsection{Motion of counterweight}
The position of the counterweight is
\begin{eqnarray}\nonumber
	\rM_M=H\eM_y-
						L_2(\sin\ta\eM_x-\cos\ta\eM_y)+
						L_3(\sin\psi\eM_x-\cos\psi\eM_y),
\end{eqnarray}
where~$\ta$ and~$\psi$ are related to the measured angles~$\te$ and~$\psi_e$ 
by~\eref{eq:theta_e} and~\eref{eq:psi_e}, 
and~$\eM_x$ and~$\eM_y$ are unit vectors shown in~\fref{fig:Trebuchet}b. 
Apart from the two measured angles, the position~$\rM_M$ depends only on the
height of the fulcrum~$H$ and the lengths of the short beam section~$L_2$ and 
the arm for the counterweight~$L_3$.

\begin{figure}[htb]
	\centering	
	\includegraphics[width=0.75\textwidth]{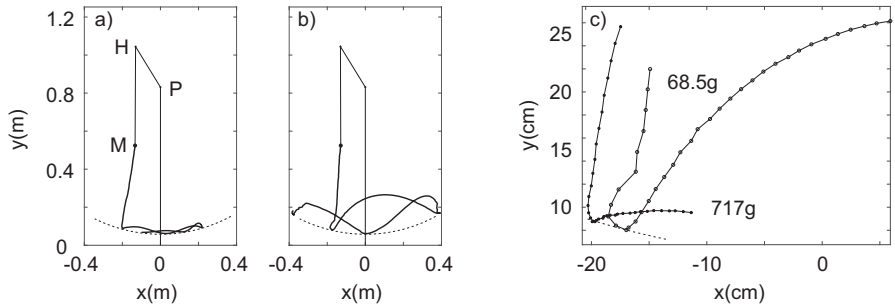}
		\caption{\label{fig:CW_motion}
		Motion of counterweight.
		a)~$m=717$g.
		b)~$m=68.5$g.
		c)~Details.}
\end{figure}

The path of the counterweight is illustrated in~\fref{fig:CW_motion} 
%
for the two projectile masses in~\fref{fig:Prim_data}. 
\Fref{fig:CW_motion}a and~\ref{fig:CW_motion}b show that 
the counterweight falls steeply from the initial position at~M until the 
fall is suddenly stopped and an oscillatory motion back and forth begins.
In~\fref{fig:CW_motion}a, the trajectory of this final motion lies quite 
close to a circle of radius~$L_2+L_3$ centered at the pivot~P and the 
circle is touched every time the configuration of counterweight and beam is 
stretched out.
This motion is much less violent than the motion shown in~\fref{fig:CW_motion}b 
with the lighter projectile where large excursions are seen both sideways and 
up down.

The transition from fall to oscillations is shown in more detail 
in~\fref{fig:CW_motion}c.
Neighboring points are separated in time by~10ms so the density of points 
reflects speed.
This is seen to be reduced rather significantly at the transition with the 
heavy projectile, but it is almost constant with the light one, and for this it
is even possible, with good will, to see a resemblance to elastic reflection 
off the circle including equal angles of incidence and reflection.
This is the theoretical behavior expected for an infinitely heavy counterweight:
Like a ball that falls from rest over a hemispherical bowl and bounces back 
elastically, the counterweight first drops vertically and thereafter goes into 
a series of free falls interrupted by elastic reflections.
\subsection{Motion of projectile and estimates of ranges, efficiencies and 
						merits}\label{sec:REEC}
\begin{figure}[htb]
	\centering	
	\includegraphics[width=0.85\textwidth]{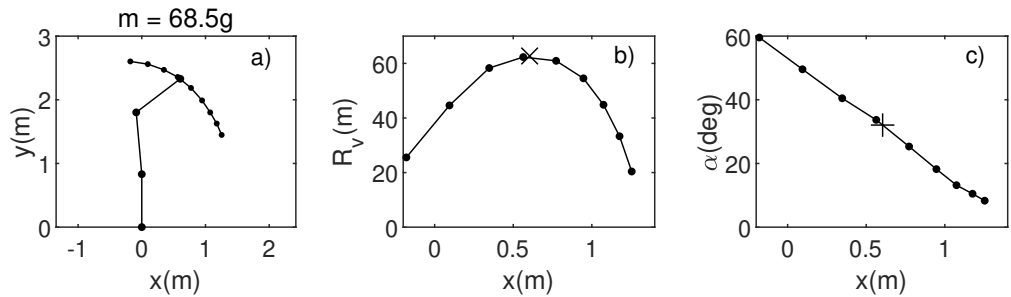}
		\caption{\label{fig:L_Range}
		a)~Path of light projectile in sling.
		b)~Range in vacuum.
		c)~Spigot angle.}
\end{figure}

The position of the projectile has the same form as the position of the counterweight
\begin{eqnarray}\label{eq:rP}
	\rM_m=H\eM_y+
						L_1(\sin\ta\eM_x-\cos\ta\eM_y)-
						L_4(\cos\phi\eM_x+\sin\phi\eM_y).
\end{eqnarray}
The functions~$\rM_m$ and~$\vM_m=\dot\rM_m$ may be taken as initial 
conditions for ballistic trajectories in vacuum.
A range function, that depends on time or position of release, 
can be extracted from these, see~\ref{app:VacRan}.
For the light projectile in~\fref{fig:Prim_data} and other parameters 
as in~\tref{tab:Param}, we find the results shown in \fref{fig:L_Range}. 
The projectile's trajectory near the position for longest range is 
plotted in~\fref{fig:L_Range}a, and the configuration of beam and 
sling at this position is shown.
%
Note that the path is not perpendicular to the sling because the spigot 
moves to the left.
\Fref{fig:L_Range}b shows calculated vacuum ranges~$R_v$ plotted as a 
function of horizontal release position.
There is a clear maximum of~$\simeq62$m near~$x=0.6$m.
Spigot angles~$\al$ are shown in~\fref{fig:L_Range}c where it is seen 
that a setting close to~$32^\circ$ leads to maximum range.

\begin{figure}[htb]
	\centering	
	\includegraphics[width=0.85\textwidth]{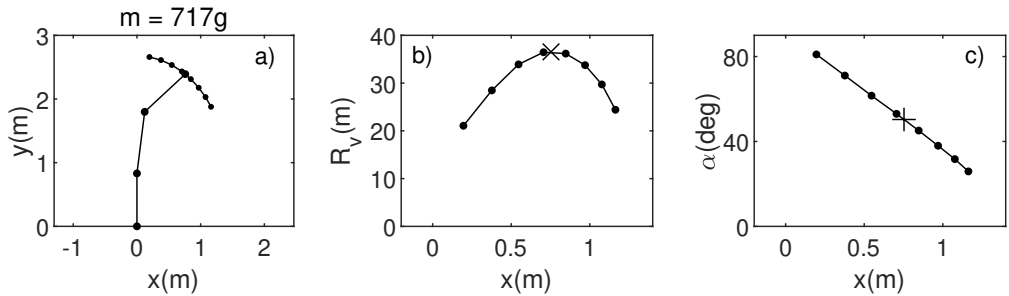}
		\caption{\label{fig:H_Range}
		a)~Path of heavy projectile in sling.
		b)~Range in vacuum.
		c)~Spigot angle.}
\end{figure}

Similar results for the heavy projectile in~\fref{fig:Prim_data} is shown 
in~\fref{fig:H_Range}.
The maximum range in vacuum is reduced to~$\simeq36$m and is found 
near~$x=0.75$m for~$\al\simeq50^\circ$.
Note that~$\ta<\pi$ at the critical configuration while~$\ta>\pi$ in
\fref{fig:L_Range}a.
%

Numerical results for the longest vacuum ranges in the two cases are collected 
in \tref{tab:Ranges}.
The ranges are given first next to the projectile masses, and then the best 
release times~$t_r$ and spigot angles~$\al$.
The initial values for the ballistic paths are positions~$(x_i,y_i)$, speeds~$v_i$ 
and climbs~$A$.
The ends of the trajectories are at~$(-R_v,0)$ where the speed in vacuum is~$v_f$ 
and the kinetic energy~$T_v=\frac{1}{2}mv_f^2$.
The values of~$T_v$ should be compared to the invested energy~$\D U=204$J.
The light projectile is seen to carry only~11\% of~$\D U$ while 
almost~70\% goes to the heavy.
The higher efficiency~$\en$ for the heavy projectile was noted already in the
discussion of the results in~\fref{fig:Prim_data} and~\ref{fig:CW_motion}.
\begin{table}[tbh]\footnotesize   
	\centering
		\begin{tabular}{cc|cc|cccc||ccc|cc}
			 $m$	& $R_v$	& $t_r$	& $\al$					& $x_i$	& $y_i$ & $v_i$	& $A$						& $v_f$ & $T_v$	& $\en$ & $\M$				& $\M_\en$		\\
			 g 		& m			& s 		&								& m 		& m 		& m/s		&  				  		& m/s 	& J			&  \%		& m$\cdot$kJ	& m$\cdot$kJ	\\\hline
			 68.5 & 62.4	& 0.533 & 32.2$^\circ$	& 0.60	& 2.33  & 25.1	& 34.4$^\circ$	& 26.0  & 23.2	& 11  	& 1.45				& 0.16 				\\
			 717	& 36.6	& 0.593 & 50.0$^\circ$ 	& 0.76	& 2.39  & 18.6	& 39.5$^\circ$	& 19.9  & 141 	& 69  	& 5.18 				& 3.57
		\end{tabular}
	\caption{Parameters for longest vacuum ranges~$R_v$ with two projectile 
						masses~$m$.
						Speed~$v_f$ and kinetic energy~$T_v$ at target.
						Efficiency~$\en$.
						Merits~$\M$ and~$\M_\en$.}
	\label{tab:Ranges}
\end{table}

Special emphasis has been placed on maximum vacuum range, but
other parameters could be considered.
If both range and damage at target are essential,
then the parameter~$\M=R_vT_v$ is a more appropriate variable,
and the efficiency could also be brought into play as in~$\M_\en=\en R_vT_v$,
although~$\en\propto T_v$ with fixed dimensions as here.
We shall refer to such functions as merits of the trebuchet.
According to \tref{tab:Ranges}, the heavy projectile outperforms the light
by both standards in spite of the shorter range.
The given merits are evaluated at the release point for maximum range,
and this probably does not result in the largest possible values.
We return to the question of optimization in~\sref{sec:VarProjMass}.
\section{Results and discussion}
\subsection{Varying projectile mass}\label{sec:VarProjMass}
\begin{figure}[htb]
	\centering	
	\includegraphics[width=1\textwidth]{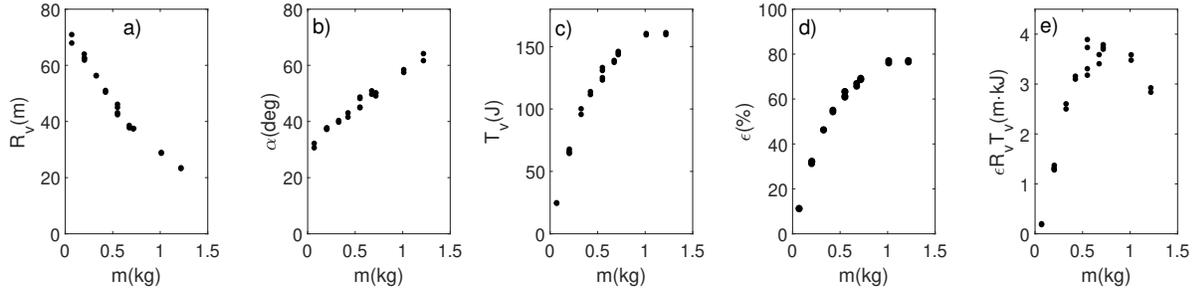}
		\caption{\label{Var_vs_m_1}
		As functions of projectile mass~$m$:
		a)~Maximum vacuum range,~$R_v$.
		b)~Spigot angle,~$\al$ in degrees. 
		c)~Energy at target,~$T_v$. 
		d)~Efficiency,~$\en$.
		e)~Merit function,~$\M_\en=\en R_vT_v$.
		Repetitions at each~$m$ expose statistical errors.}
\end{figure}

Standard tennis and petanque balls supplemented by natural stones with almost 
spherical shapes and different masses were used as projectiles in a series 
of measurements with the constant parameters listed in~\tref{tab:Param}.
Maximum vacuum ranges~$R_v$ and required spigot angles~$\al$ were derived 
for each projectile as discussed in~\sref{sec:REEC}, and
the results are plotted in~\fref{Var_vs_m_1}.
%
They show a quite significant decrease of~$R_v$ as a function of~$m$ 
and a spigot angle~$\al$ that increases such that the heavier projectiles
swing further relative to the beam than the lighter.
The kinetic energy at target~$T_v$ was also derived and it shows a strong increase
that seems to level off or reach a maximum. 
The efficiency is likewise seen to be an increasing function of~$m$, 
and this is taken into account in the merit function~$\M_\en=\en R_vT_v$, 
introduced in~\sref{sec:REEC}. 
$\M_\en$ shows a clear maximum close to~$m=0.7$kg, which is not far from the heavy 
mass in~\tref{tab:Ranges}.

\begin{figure}[htb]
	\centering	
	\includegraphics[width=1.0\textwidth]{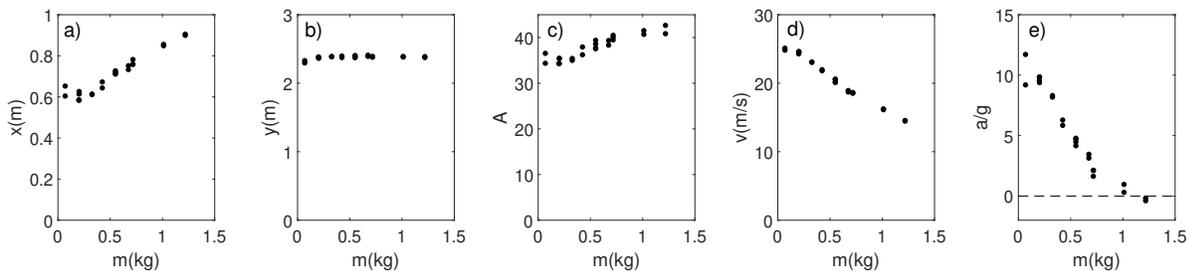}
		\caption{\label{fig:Var_vs_m_2}
		Initial conditions for maximum vacuum range \textit{vs} projectile mass.
		a+b) Position of projectile at release~$(x,y)$.
		c) Initial climb~$A$ in degrees.
		d) Speed at release~$v$.
		e) Acceleration along path at release~$a=\dot v$ in units of gravity~$g$.}
\end{figure}

The initial conditions for the longest ballistic trajectories are shown 
in~\fref{fig:Var_vs_m_2}.
The variable~$x$ in~\fref{fig:Var_vs_m_2}a is the horizontal position of 
the projectile at release.
%
The target is found at a negative value of~$x$, so a positive release position 
subtracts from the range.
The variable~$y$ in~\fref{fig:Var_vs_m_2}b is the height at start measured from the 
base of the trestle, and
it is added almost unchanged to the range for a target at ground level.
The initial climb of the projectile~$A$ in~\fref{fig:Var_vs_m_2}c is a little 
below~45$^\circ$ but approaches this value at large~$m$.
The speed at release~$v=|\vM|$, which is the initial speed in the ballistic 
projectile path, is shown in~\fref{fig:Var_vs_m_2}d.
It decreases with~$m$ and reflects the same trend for the range~$R_v$.
The acceleration~$\aM$ projected on the velocity at release~$a=\aM\cdot\vM/v$
is shown in~\fref{fig:Var_vs_m_2}e.
It is quite large and positive for light projectiles so these are still 
gaining speed at release for longest range. 
The range maximizes at an initial climb~$A$ of~$45^\circ$ for constant speed,
but since speed is increasing for decreasing~$A$, a smaller~$A$ is advantageous
and~\fref{fig:Var_vs_m_2}c shows that~$A\simeq35^\circ$ is best for the lightest 
projectiles.

The merit values~$\M$ in~\tref{tab:Ranges} are not maximized, so another
release time may increase their values, and this is indeed the case for
light projectiles.
If release takes place a little later~$\D t$ than specified in~\tref{tab:Ranges}, 
the range decreases slowly like~$1-(\D t/t_R)^2$, but the energy increases linearly
like~$1+\D t/t_E$ such that the product increases.
Here,~$t_R$ depends on the curvature~$d^2R/dx^2$ at maximum and the speed~$v_x=v\cos(A)$, 
and~$t_E$ on the acceleration~$a$.
The curvature can be read from~\fref{fig:L_Range} and~$v$,~$A$ and~$a$ from~\fref{fig:Var_vs_m_2}.
We thus find~$t_R\simeq45$ms and~$t_R\simeq110$ms, which leads to a maximum of~$\M$ 
at~$\D t\simeq10$ms (the time between measuring points). 
Here~$\M$ has increased by~$\simeq4\%$, the range~$R$ is reduced by~$\simeq4\%$ and the kinetic 
energy~$T$ is larger by~$\simeq9\%$.
This also improves the efficiency by~$\simeq9\%$ from~$\simeq11\%$ to~$\simeq12\%$.
For the heavy projectiles with~$a/g\simeq0$, range and speed maximize almost at 
the same time, so~$\M$ is already at maximum.
Optimization is discussed in detail in~\cite{ref:Horsdal}.

\begin{figure}[htb]
	\centering	
	\includegraphics[width=1\textwidth]{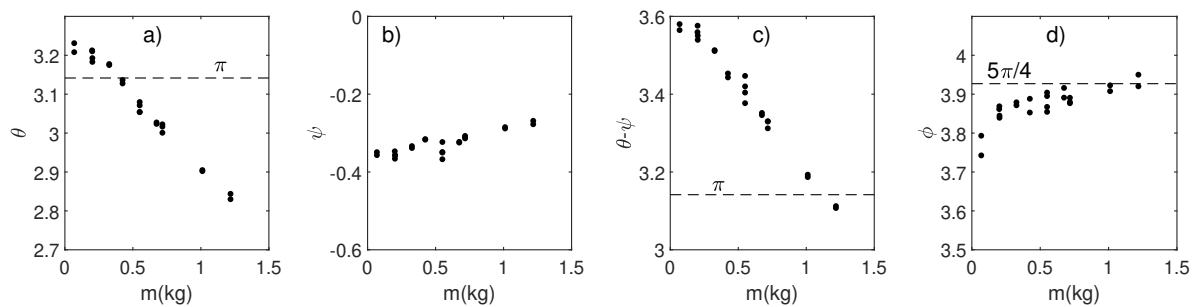}
		\caption{\label{Angles_Release}
		Angular coordinates at release for maximum vacuum range
		\textit{vs} projectile mass~$m$.
		Same angular ranges of 0.6 radians.
		a)~Beam,~$\ta$.
		b)~Counterweight,~$\psi$.
		c)~Beam and counterweight,~$\ta-\psi$.
		d)~Sling,~$\phi$.}
\end{figure}

\Fref{Angles_Release} shows the configuration of the trebuchet at release 
expressed by the angular coordinates for beam, counterweight and sling.
The beam is close to vertical at~$\ta=\pi$ for all 
projectiles,~\fref{Angles_Release}a, 
but while it has just passed this point for the lightest it has not come so 
far for the heaviest.
This was illustrated already in~\fref{fig:L_Range} and~\ref{fig:H_Range}.
%
The counterweight angle~$\psi$ is small and negative at 
release,~\fref{Angles_Release}b,
but during a shot it starts from zero and is first positive before it becomes 
negative.
The angle between beam and counterweight,~$\ta-\psi$, is shown 
in~\fref{Angles_Release}c.
The configuration is stretched out when~$\ta-\psi=\pi$, and this happens 
near~$m=1$kg or close to maximum merit and an initial climb of~$A=45^\circ$.
The sling angle~$\phi$ shown in~\fref{Angles_Release}d is near~$5\pi/4$ at 
release, or~$\pi/4=45^\circ$ over the horizon as seen also 
in~\fref{fig:L_Range} and~\ref{fig:H_Range}. 
\newpage
\subsection{Varying counterweight}
\begin{figure}[htb]
	\centering	
	\includegraphics[width=0.70\textwidth]{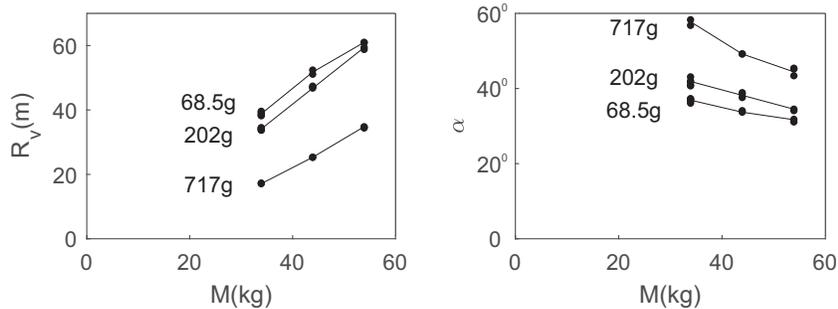}
		\caption{\label{fig:Range_vs_CW}
		Maximum vacuum ranges~$R_v$ and spigot angles~$\al$ 
		\textit{vs} counterweight~$M$.
		Projectile masses next to curves.}
\end{figure}

Two smaller masses of the counterweight and three projectile masses were used in 
a series of experiments to illustrate the dependence on counterweight mass~$M$.
The other parameters of the trebuchet are listed in \tref{tab:Param}.
\Fref{fig:Range_vs_CW} 
%
shows the longest vacuum range~$R_v$ and the required 
spigot angles~$\al$.
A lighter counterweight clearly leads to a significant reduction of 
range irrespective of projectile mass.

%
\subsection{Varying sling length}
\begin{figure}[htb]
	\centering	
	\includegraphics[width=0.70\textwidth]{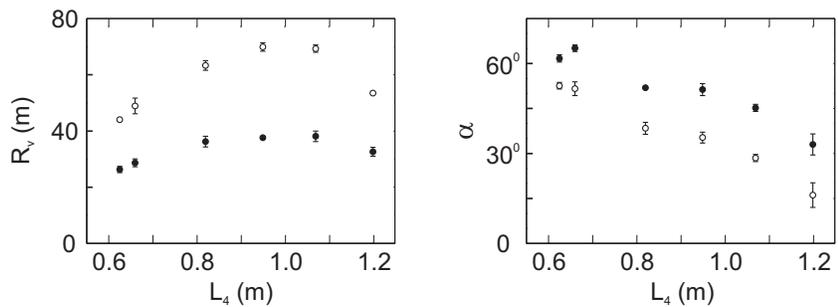}
		\caption{\label{fig:Philip_I}
		Maximum vacuum ranges~$R_v$ and spigot angles~$\al$ 
		\textit{vs} sling lengths~$L_4$.
		Open points:~$m=170$g.
		Full points:~$m=700$g.}
\end{figure}

Maximum vacuum ranges~$R_v$ and corresponding spigot angles~$\al$ were 
determined also for a number of sling lengths~$L_4$ and two projectile 
masses~$m$.
Other parameters of the trebuchet are listed in~\tref{tab:Param}.
The ranges~$R_v$ for the two masses are shown in~\fref{fig:Philip_I} and
%
each shows a broad maximum located close to the sling length~$L_4$ of~0.87m
used in the previous measurements.
The spigot angle~$\al$ varies significantly with~$L_4$ such that a longer~$L_4$
implies a smaller~$\al$ and vice versa.
\subsection{Mechanical energy}
The potential energy of the trebuchet is assumed to be zero when the 
counterweight is at rest in its lowest position and a projectile lies ready in 
the trough at ground level.
The initial mechanical energy prior to a shot is then the potential 
energy~$\D U$ in~\eref{eq:DU} added during loading, and the mechanical energy is 
thereafter either constant or strictly decreasing because of irreversible losses.

The mechanical energy is the sum of all kinetic and potential energies.
For the degrees of freedom in~\fref{fig:Trebuchet}, the total kinetic energy is
\begin{eqnarray}
	T		=\frac{1}{2}m\dot\rM_m^2
			+\frac{1}{2}\I_m^{cm}\dot\phi^2
			+\frac{1}{2}M\dot\rM_M^2
			+\frac{1}{2}\I_M^{cm}\dot\psi^2
			+\frac{1}{2}\I_b\dot\ta^2.
	\label{eq:T}
\end{eqnarray}
The first two terms of the sum is the mechanical energy of the projectile.
It has translational and rotational speeds~$|\dot\rM_m|$ and~$\dot\phi$, 
respectively, and~$\I_m^{cm}$ is the moment of inertia with respect to 
the center of mass.
The next two terms constitute the mechanical energy of the counterweight,
and the last term is the mechanical energy of the beam whose moment of 
inertia relative to the pivoting axle is~$\I_b$.
The two terms that represent the energies of rotation for projectile and 
counterweight contribute only little.

The potential energy for the same degrees of freedom is 
\begin{eqnarray}
	U=mgy_m+Mg(y_M-y_M^f)
			+m_bg(y_{m_b}-y_{m_b}^f).
	\label{eq:U}
\end{eqnarray}
Here~$y_m$,~$y_M$ and~$y_{m_b}$ are the instantaneous heights above ground 
of the centers of mass for projectile, counterweight, and beam, respectively.
The final values at rest after a shot are~$y_m^f=0$,~$y_M^f=H-(L_2+L_3)$ 
and~$y_{m_b}^f=H+L_{CM}$.

The motion of sling (and projectile) is measured until the time when the 
pin~P in~\fref{fig:SlyngeSpigot}a is stopped by the beam, but the
projectile remains in the pouch a little longer.
The motion of sling and projectile was calculated in this short interval 
of time by treating the system as a pendulum in an accelerated coordinate
system that follows the spigot. 
The initial conditions for the motion was determined by the sling 
motion leading up to the pin stop, and it was followed until the direction 
of the projectile velocity dips below~$-45^\circ$ relative to the horizon, 
such that a point just in front of the engine is hit.
The extrapolation of the motion is short, but important, because 
mechanical energy is exchanged also here, and it allows all kinetic 
and potential energies in~\eref{eq:T} 
or~\eref{eq:U} to be calculated at all times. 

\begin{figure}[htb]
	\centering	
	\includegraphics[width=0.95\textwidth]{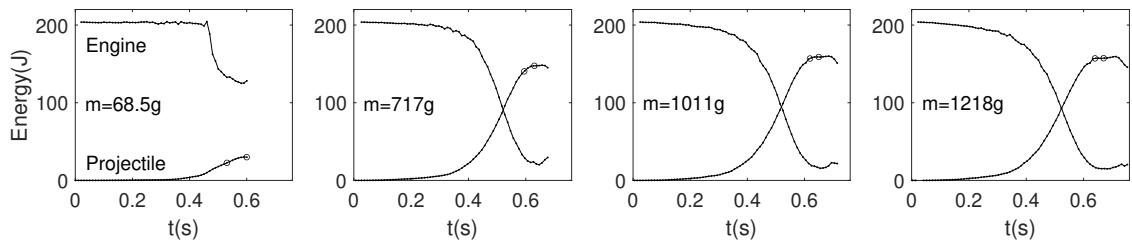}
		\caption{
		Mechanical energies~\textit{vs.} time until release
		for four projectile masses~$m$.}
		\label{fig:Loss_Energies_1}
\end{figure}

The total mechanical energies of engine (counterweight and beam) 
and projectile are shown as functions of time until release 
in~\fref{fig:Loss_Energies_1}
%
for the lightest and the three heaviest projectiles used in the experiments.
The release time for longest range followed by the time when the pin is stopped
are marked by circles.
The energy of the engine is~$\D U=204$J at start and a small fraction of this
is subsequently transferred to the light projectile whereas the heavier 
absorb much more. 
The process is also slower for the heavy projectiles, and the extrapolations 
from pin stop to actual release are longer.
When the projectile leaves the trebuchet, it carries the mechanical energy 
plotted at the end of the time series.
We do not treat this as a lost energy, but include it as a constant contribution 
to the total energy after release.

\begin{figure}[htb]
	\centering	
	\includegraphics[width=0.95\textwidth]{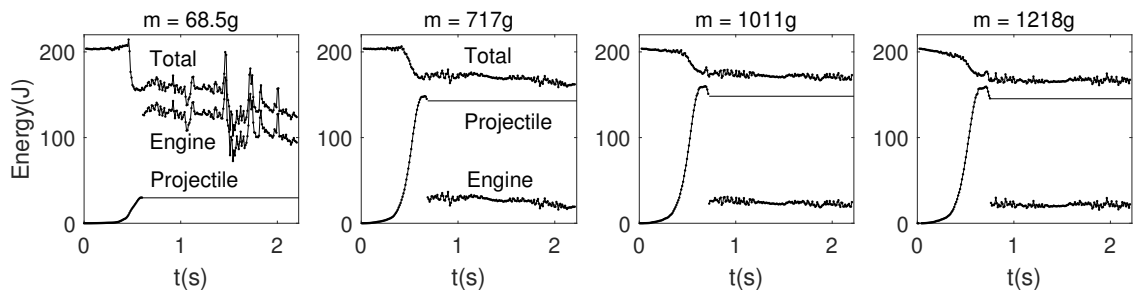}
		\caption{
		Mechanical energies covering full durations of measurements.}
		\label{fig:Loss_Energies_2}
\end{figure}

\Fref{fig:Loss_Energies_2} shows mechanical energies over the full duration of
a measurement with the constant projectile contribution after release represented 
by horizontal lines.
The total energy shows increases over short intervals of time at each projectile 
mass but most clearly for the lightest.
%
These increases are reproducible and reveal systematic errors.
Most are seen after the projectile is released which incriminates the movement
of beam or counterweight.
The errors are largest with the light projectile for which the inner movement 
of the engine is quite violent, 
and even more so for blank shots where the trestle clearly tilts and moves 
unless extra weight is added, most often by a person standing on the base.
The data for the heavier projectiles are less suspicious, but still show increases.
We therefore look apart from the long term data for the light projectile, 
and suspect that one or more degrees of freedom, that absorb and release 
mechanical energy, are missing in~\eref{eq:T} and~\eref{eq:U}.
This could be elastic bending of the beam or a small, undetected movement of the 
fulcrum due to a flexible trestle.
The last possibility was examined by a simple model in which the bearings for 
the pivoting axle are fixed to a heavy mass that is free to slide without 
friction on a horizontal surface.
When the heavy mass is adjusted to limit the movement to a fraction of a centimeter, 
we see a few joules of energy being exchanged after about one second so this could 
be an explanation, but we can not go any further without actual knowledge of the 
motion, which we do not have.
The model also indicates very little influence on range or energy at target, 
even for strong movement of the bearings.
This is because strong horizontal forces are seen only long after the projectile is 
launched and the counterweight has begun its final, oscillatory motion.

In addition to systematic errors, the data in~\fref{fig:Loss_Energies_1}
and~\ref{fig:Loss_Energies_2} also suffer from random errors that 
look like noise. 
This can be traced to the counterweight's kinetic energy, which depends 
sensitively on~$\dot\psi$ and~$\dot\ta$, but the noise is clearly generated mostly 
by fluctuations of~$\dot\psi$.
The angle~$\ta$ therefore seems to be measured more precisely than~$\psi$.
\subsection{Loss of mechanical energy to friction and air resistance}
\begin{figure}[htb]
	\centering	
	\includegraphics[width=1\textwidth]{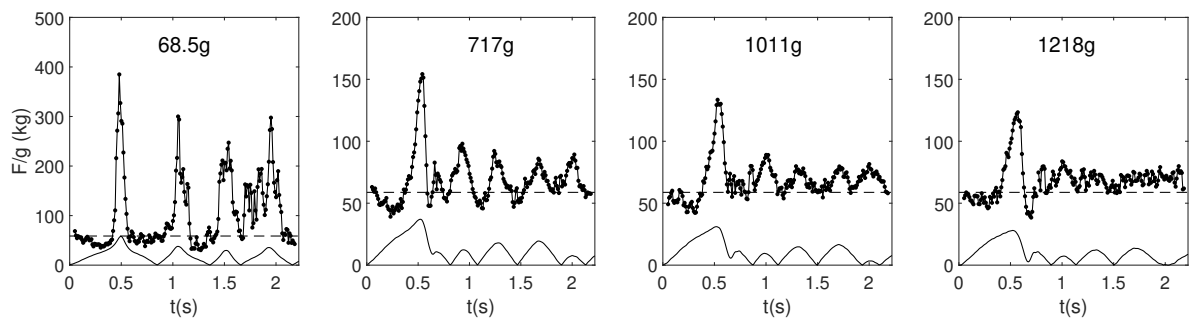}
		\caption{
		Full points:
		Reaction force~$|\FM_P|$ at pivot in units of mass for four projectile masses.
		Broken line: Final value~$M+m_b$.
		Full curve: $|\dot\ta|$ in arbitrary units.}
		\label{fig:Forces} 
\end{figure}

Mechanical energy is inevitably lost to friction and air resistance.
Friction is found at the bearings for the pivoting axle of the beam where wood
moves against wood, and at the hinge for the counterweight where the materials are
stainless steel.
Air resistance primarily affects the sling motion because of the relatively 
large aerodynamic cross section of the pouch and the high speed shortly before release 
of the projectile.
The steep fall of total mechanical energy seen near~0.5s 
in~\fref{fig:Loss_Energies_2} is thus tentatively attributed to heat generation when 
the initial fall of the counterweight is suddenly interrupted, but also to turbulence 
from fast sling motion although this is less important as will be seen. 
The magnitudes of the reaction forces at the pivot for the beam~$|\FM_P|$ 
and the hinge for the counterweight~$|\FM_H|$ are then large and so are the respective 
sliding speeds~$R_P\dot\ta$ and~$R_H(\dot\ta-\dot\psi)$, 
where~$R_P$ and~$R_H$ are radii.
These factors determine the rate of heat generation as estimated by the 
standard model for friction
\begin{eqnarray}\label{eq:Power}
	P_f=\mu_p|\FM_P|R_P|\dot\ta|+\mu_H|\FM_H|R_H|\dot\ta-\dot\psi|,
\end{eqnarray}
where~$\mu_p$ and~$\mu_H$ are empirical friction coefficients.
The reaction force~$\FM_P$ balances all other forces that act on the bearings for
the beam treated as a rigid body.
These are the forces at the hinge~$\FM_H$, the center of mass for the beam~$\FM_{CM}$,
and the spigot~$\FM_S$, so we have
\begin{eqnarray}\nonumber
	\FM_P	&=\FM_H+\FM_{CM}+\FM_S 						\\ \nonumber
				&=M(\ddot\rM_M+g\eM_y)+m_b(\ddot\rM_{CM}+g\eM_y)+m(\ddot\rM_m+g\eM_y).
\end{eqnarray}
The force $|\FM_H|$ is always much larger than~$|\FM_{CM}|$ or~$|\FM_S|$, 
so~$|\FM_P|\simeq |\FM_H|$ throughout a shot.
$|\FM_P|$ is shown in~\fref{fig:Forces} after division by~$g$ to make the force appear 
as a gravitational weight.
\vspace{2pt}
%
For each projectile mass,~$|\FM_P|/g$ reaches a global maximum after about~0.5s when 
the initial fall of the counterweight is stopped, and the angular speed~$|\dot\ta|$
also has a maximum here.
The maximum of~$|\FM_P|/g$ is quite high with the light projectile for which it almost 
reaches~$400$kg, and the secondary maxima that follow go as high as~$300$kg.
With the heavier projectiles, the global maxima are more than a factor of two smaller 
and the secondary maxima are progressively reduced such that they have almost 
disappeared at~$m=1218$g, where~$\FM_P/g$ is close to the final value
of~$M+m_b=58.76$kg already after one second. 
The maxima of the angular speeds~$|\dot\ta|$ almost coincide with the maxima
of~$|\FM_P|/g$. 

\begin{figure}[htb]
	\centering	
	\includegraphics[width=1\textwidth]{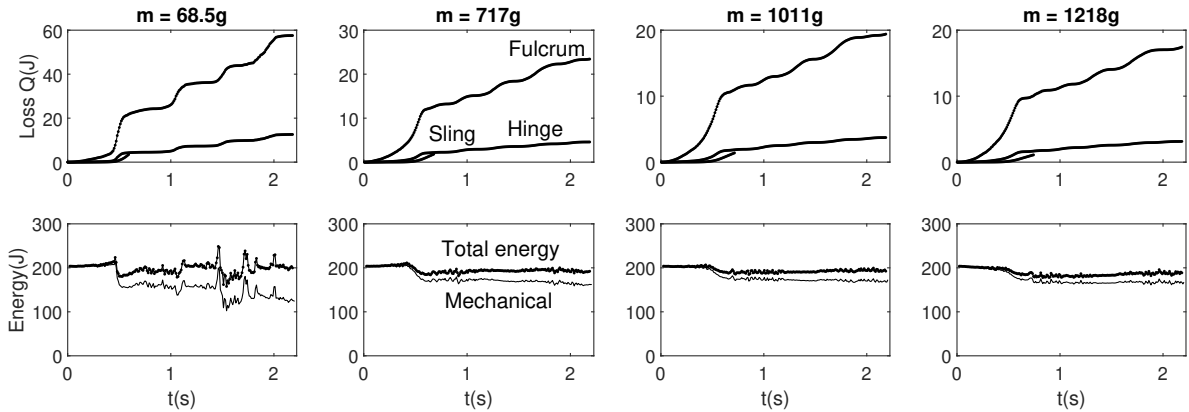}
		\caption{
		Upper panel:
		Frictional losses at fulcrum for beam and hinge for counterweight. 
		Aerodynamic loss for sling.						
		Friction coefficients:~$\mu_f=\mu_h=0.35$.
		Radii:~$R_P=1.7$cm and~$R_H=0.25$cm.
		Aerodynamic constant~$C=0.25$,~$\rho_d=1.225$kg/m$^3$ and~$A_P=0.01$m$^2$,
		see~\ref{app:AirD}.
		Lower panel: 
		Total mechanical energy as in \fref{fig:Loss_Energies_2}
		and total energy including losses.}
		\label{fig:Loss_Energies_3} 
\end{figure}

The accumulated loss of mechanical energy due to friction at time~$t$ is
\begin{eqnarray}\nonumber
	Q_f=\int_0^{t}P_f(t')dt',
\end{eqnarray}
with~$P_f$ from~\eref{eq:Power}, 
and the aerodynamic loss from the sling motion is 
\begin{eqnarray}\nonumber
	Q_a=\frac{1}{2}C\rho_dA_P\int_0^{t}v_m^3(t')dt',
\end{eqnarray}
where~\eref{eq:AeroForce} is used.
The two terms in the frictional loss~$Q_f$ and the aerodynamic 
contribution~$Q_a$ are shown separately in the upper panel
of~\fref{fig:Loss_Energies_3}.
%
The loss at the fulcrum dominates the frictional losses because the sliding
speed is larger~$R_P\dot\ta\gg R_H(\dot\ta-\dot\psi)$ and the reaction forces are 
almost equal. 
The aerodynamic component is first quite small, but finally rises to approach
the loss at the hinge.

The lower panel in~\fref{fig:Loss_Energies_3} shows the sum of total mechanical
energies and calculated losses.
Energy is conserved, so this sum should equal the initial energy~$\D U$ at all times.
The sum is seen to be a little low, but the deviations for the heavier projectiles are 
quite small and possibly due to missing degrees of freedom as discussed earlier.
\subsection{Semi-empirical ranges. Ballistics in air}
The experimental data discussed so far were obtained in a laboratory with 
sufficient distances to ceiling and walls to safely accommodate sling and 
projectile throughout an entire shot.
The projectiles reached maximum heights~$\simeq3$m, and were shot into and
captured by a tightly packed pile of straw just in front of the trebuchet.
The vacuum ranges,~$R_v$, derived from the measurements were corrected for 
aerodynamic losses along the ballistic path to give semi-empirical calculated 
ranges in air,~$R_a$.
The expressions used for vacuum ranges and losses are given in~\ref{app:VacRan}
and~\ref{app:AirD}.
\subsection{Field measurements of range}
\begin{figure}[htb]
	\centering	
	\includegraphics[width=0.30\textwidth]{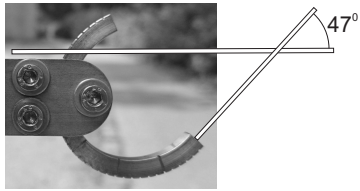}
		\caption{\label{fig:Spigot_Setting}
		Semicircular spigot clamped to beam by three bolts.
		Spigot angle shown.}
\end{figure}

To test these semi-empirical ranges, the trebuchet was taken out 
to a flat field on warm days with little wind (close to standard temperature 
and pressure) and fired with two projectile masses~$m$ and two lengths of the 
sling~$L_4$.
Many shots with different settings of the spigot were required in 
each case to find the longest field range~$R_f$, and this effort required
the work of at least two people.
One to load and fire the trebuchet, and another to observe and log where 
the projectile hits ground.
The spigot has markings for easy and reproducible setting, and is shown 
in~\fref{fig:Spigot_Setting}.
The force that act on it with a heavy projectile rises briefly to~$\simeq250$N. 
This necessitates hardening after machining to prevent bending.
%

Semi-empirical predictions from the present experiments, field measurements and
purely theoretical values are listed in~\tref{tab:Examples_I}.
The random errors of the semi-empirical values~$R_v$ are calculated from
identical, repeated shots, and the larger uncertainty for~$R_a$ comes from 
an empirical, aerodynamic constant~$C$, see~\ref{app:AirD}.
\begin{table}[htb]\footnotesize   
	\centering
		\begin{tabular}{c|cc|ccc|cc}
											&$L_4$& $m$		& $R_v$ 				& $R_a$				& $\al_v$ 			& $R_f$ 				& $\al_f$ 			\\
											&m 		& g			& m							& m						&   						& m  						&   						\\\hline	
			Semi-empirical	&0.87	&	 717	& 37.5$\pm$0.1	&35.1$\pm$0.5	&	50.0$^\circ$	& 							&								\\	
			predictions			&0.87	&	1011	& 29.0$\pm$0.2	&27.7$\pm$0.5	&	58.5$^\circ$	& 							&								\\\hline
			Field 					&0.87	&	700		&								&							&								& 34.4 $\pm$1.5	&	36.2$^\circ$	\\
			measurements		&0.87	&	1011 	&								&							&								&	24.9$\pm$1.5	&	32.5$^\circ$	\\
											&0.93	&	700  	& 							&							&								& 37.5$\pm$1.5	&	19.7$^\circ$	\\\hline
			Theoretical 		&0.87	&	717		&	42.8					&	39.5				&								& 							&								\\
			values					&0.87	&	1011 	&	32.2 					&	30.6				&								&								&
											
		\end{tabular}
	\caption{
	Semi-empirical maximum ranges in vacuum~$R_v$ and in air~$R_a$.	 
	Ranges measured in the field~$R_f$.
	Ab initio theoretical values of~$R_v$ and~$R_a$.
	Predicted spigot angles~$\al_v$ and field values~$\al_f$.
	Stones of mass density~$\rho_p=3$g/cm$^3$ and~$C=0.5$.
	The design parameters are those listed in \tref{tab:Param} except for the 
	sling length~$L_4$ of~0.93m used in the field.}
	\label{tab:Examples_I}
\end{table}
The random errors for the field measurements~$R_f$ are based on several 
repetitions under the same experimental conditions as far as possible.

The field values~$R_f$ are seen to agree well within uncertainties with the 
semi-empirical predictions~$R_a$
(when disregarding a small difference of~17g in projectile mass, see~\fref{Var_vs_m_1}a).
The field data with the increased sling length of~0.93m also shows a longer range.
This is in good agreement with the data in \fref{fig:Philip_I}, which indicates 
a best sling length close to one meter.

The theoretical ranges in~\tref{tab:Examples_I} were found by the use 
of an internet facility~\cite{ref:VTreb} and by independent integrations.
The ranges in air are several~($\simeq4$) standard deviations longer 
than the field values.
The calculations, however, disregard internal friction and aerodynamic 
losses during the acceleration of the projectile prior to its release.
\Fref{fig:Loss_Energies_3} indicates a loss of mechanical energy at release for
longest range of~$\simeq5\%$ and this loss is carried mostly by the projectile.
Range and kinetic energy at release are approximately proportional,
so~$\D R/R\simeq\D T/T\simeq5\%$.
This estimate suggests that the inclusion of losses will lower the discrepancy
significantly.

The predicted spigot settings~$\al_v$ in~\tref{tab:Examples_I} are much larger than 
the ones used in the field~$\al_f$.
We have assumed that the ring slides off the semicircular spigot immediately when 
the sling direction becomes perpendicular to the direction of the spigot.
This seems to be a reasonable assumption because the ring glides easily on the
spigot and is therefore close to the edge at the critical time.
Ring, cord and pouch, however, all have some inertia so it takes a little extra 
time for the projectile to free itself from the pouch.
This time is estimated by~$\D t=\D\phi/\dot\phi$, where~$\D\phi\simeq1/3$rad is 
taken from the differences of spigot angles in~\tref{tab:Examples_I} 
and~$\dot\phi\simeq15$rad/s from~\fref{fig:Var_vs_m_2}d 
(spigot almost at rest, so~$v\simeq L_4\dot\phi$).
We find~$\D t\simeq20$ms which is short indeed and a likely explanation for the 
discrepancy, but difficult to qualify further.
\section{Optimized design}
The experimental trebuchet was built on intuition and a desire to have a durable 
and safe engine.
Theoretical optimization of lengths and masses was not attempted, but the sling length 
and projectile mass that gives the best product of range and kinetic energy at target
was found experimentally for constant values of remaining parameters.
The result is shown in the first row of~\tref{tab:Opt}.
\begin{table}[htb]\footnotesize
	\centering
		\begin{tabular}{cc|cccc|cc|ccc|cccc}
			 $R$	& $T$	& $L_1$	& $L_2$	& $L_3$	& $L_4$	& $D$		& $H$		& $M$		& $m$		& $m_b$	& $\en$ & $S_P$	& $u_P$	& $u_{CM}$ 	\\
			m 		& J 	& cm 		& cm 		& cm 		& cm 		& cm 		& cm 		& kg 		& g 		& kg 		& \% 		& \% 		& mm 		& mm 				\\\hline
			36.6 	& 141	& 97.5	& 25.0	& 51.5	& 87.0	& 7.5		& 83.1	& 53.9	& 717		& 4.86 	& 73		& 0.03	& 0.6 	& 0.8				\\\hline
			36.6 	& 141	& 128		& 36.2	& 63.9	& 111		& 4.9		& 109		& 26.3	& 697		& 2.16 	& 91		& 0.07	& 4.3 	& 5.9				\\
			50 		& 25	& 97.2	& 21.3	& 55.0	& 88.0	& 2.8		& 82.9	& 7.88	& 95		& 0.50 	& 93		& 0.07	& 3.2 	& 5.0				\\			
			50 		& 350	& 157		& 42.7	& 80.4	& 137		& 6.4		& 134		& 55.4	& 1281	& 4.51	& 91		& 0.08	& 5.2 	& 7.4						
		\end{tabular}
	\caption{The experimental trebuchet and optimized design.
	Strain at pivot~$S$. Deformation at pivot~$u_P$ and at center of mass $u_{CM}$.}
	\label{tab:Opt}
\end{table}
The efficiency~$\en=73\%$ is the estimated ideal value for the engine without losses.
It is derived from the measured efficiency of~$\simeq70\%$ (see~\tref{tab:Ranges})
and engine losses of~$\simeq5\%$.

The three remaining rows of~\tref{tab:Opt} show optimized design obtained as described 
in~\cite{ref:Horsdal}.
However, the analytical procedures laid out there do not cover small designs like the present, so 
ab initio numerical calculations were done in each case.
The first has the same capacity as the experimental trebuchet, but it is~$\simeq25\%$ larger, 
the counterweight and the beam are~$\simeq50\%$ lighter, and the ideal engine efficiency 
has risen to~91\%.
The remaining two optimized design have a longer range of~$R=50$m 
and different kinetic energies at target~$T$.
The one with~$T=25$J has the same height~$H$ of the fulcrum as the experimental design and throws
the projectile a distance of~60 times the height, but the projectile is light.
The other with~$T=350$J has about the same counterweight mass as the experimental design and 
throws a much heavier projectile, but the distance relative to height is reduced to~37.
The optimized design have significantly improved efficiencies, and are generally much lighter
and slightly larger.

The small beam masses~$m_b$ and diameters~$D$ bring up the question of strength so we show 
explicitly in~\tref{tab:Opt} values of strain~$S$ and deformation~$u$.
When the trebuchet is loaded prior to a shot, soldiers pull the beam down with ropes attached
near the spigot. 
The bending is largest when the beam is horizontal and the quasi-static reaction force 
at the fulcrum from the trestle is then~$F_s=Mg(1+L_2/L_1)$.
The strain varies along the beam and is largest at the fulcrum where the 
curvature~$\partial^2u/\partial x^2$ is at its maximum.
Strain and curvature are related by
\begin{eqnarray}\nonumber
	S=\frac{D}{2}\frac{\partial^2u}{\partial x^2},
\quad\mathrm{and~at~the~pivot}\quad
	\M_e\I\left.\frac{\partial^2u}{\partial x^2}\right|_P=F_s\frac{L_1L_2}{L_1+L_2}
\end{eqnarray}
where~$\M_e$ is Young's modulus and~$\I=(\pi/64)D^4$ the second moment of area.
The deformations relative to a rigid beam at the pivot and at the center of mass
or middle of the beam are, respectively,
\begin{eqnarray}\nonumber
	u_P = \frac{1}{3}\frac{F_s}{\M_e\I}\frac{L_1^2L_2^2}{L_1+L_2}
\qquad\mathrm{and}\qquad
	u_{CM} = \frac{1}{16}\frac{F_s}{\M_e\I}L_2(L_1+L_2)^2.
\end{eqnarray}
The static values given in~\tref{tab:Opt} are small and considered safe
with~$S_P\ll 1\%$.
The maximum deformation is found close to the center of mass and here
it is only a few millimeters per meter of beam length, and much smaller with
the experimental design.
The load on the beam varies during a shot and may rise briefly to
a few times~$Mg$, but the strengths remain sufficient.

\section{Summary and conclusions}
A wealth of information can be extracted from a measurement of the inner 
movement of a trebuchet during only a single shot. 
It is therefore affordable in terms of time to carry out systematic
studies of dependencies on several parameters. 
Maximum projectile range is a key variable, so it was emphasized and measured
as a function of projectile mass, length of sling, and counterweight mass.
Loss of range by friction and aerodynamics was estimated, and predicted ranges 
were compared with measured field ranges.
The kinetic and potential energies of the projectile at release are important 
too and were also derived.
The sum is the vacuum value of the mechanical energy delivered to 
a target at the level of the engine.

The measured total mechanical energy must be a decreasing function of 
time because of inevitable irreversible losses.
If it rises even briefly, a degree of freedom that absorbs and releases 
energy is overlooked or a constant like the moment of inertia for the beam is 
determined incorrectly.
The total mechanical energy is thus a diagnostic tool and was used as such to 
exclude some measurements concerning the inner movement 
of the engine long after release for small projectile masses.

Loss of mechanical energy during acceleration of the projectile by the engine
is mainly due to friction at bearings and hinges, 
but aerodynamic drag on the sling also contributes. 
All losses were determined, and when added to the derived mechanical energies, 
the resulting total energy was found to be approximately constant 
provided the projectile mass was not too small.
The losses then amount to~$\simeq5\%$ of the available mechanical energy of~$\simeq200$J,
and the efficiency at longest range is near~70\% for a projectile mass~$\simeq700$g. 
The ideal engine efficiency in the absence of losses is~$\simeq73\%$, but an
optimized design with the same capacity has~$\en=91\%$.

The acceleration of the projectile at release projected on its velocity is almost
zero at~700g, but positive for smaller masses and negative for larger.
The kinetic energy~$T$ of lighter projectiles is thus still increasing at release, 
so delaying this a little to maximize the merit function~$\M=RT$ increases the
kinetic energy at target~$T$, but reduces~$R$.
For heavier projectiles where the acceleration is negative at release so that
the projectile works back on the engine, an earlier release has the same effect.

The reaction force at the fulcrum rises briefly during a shot to values 
near three times the gravity~$Mg$ of the counterweight or even more for 
light projectiles.
The horizontal component approaches~$Mg$.  
These forces determine friction losses and required strengths of beam, axles, bearings,
and trestle.
The size of the sling force on the projectile increases briefly to~$\simeq25mg$.
Ropes, pouch, and spigot must be strong enough to withstand this load.
\newpage
\hspace{150pt}\textbf{APPENDIX}
\appendix
\section{Beam with fittings and model for moment of inertia}\label{app:Model}
The weights of the fittings that hold spigot and counterweight are represented in a
simple model for the beam by two point masses~$\tm_1$ and~$\tm_2$ placed at the 
respective distances~$\tL_1$ and~$\tL_2$ from the pivot.
These four parameters are related to measured parameters by
\begin{eqnarray}
	m_b\quad			&=\hm+\tm_1+\tm_2 \nonumber\\
	m_bL_{CM}\quad&=~\frac{1}{2}\hm(L_1-L_2)+\tm_1\tL_1-\tm_2\tL_2 			\nonumber\\
	\I_b					&=~\frac{1}{3}\hm\left(L_2^2-L_1L_2+L_2^2\right)+\tm_1\tL_1^2+\tm_2\tL_2^2,
	\nonumber
\end{eqnarray}
where~$\hm$ is the beam mass without fittings, 
and remaining parameters are given in~\tref{tab:Param}.
The four model parameters are collected on the left hand side of 
the matrix equation
\begin{eqnarray}\label{eq:Matrix}
	\left\{
	\begin{array}{cc}
		1				& 1						\\
		\tL_1 	& -\tL_2 			\\
		\tL_1^2 & \tL_2^2
	\end{array}
	\right\}
	\left\{
	\begin{array}{c}
		\tm_1	 \\
		\tm_2
	\end{array}
	\right\}	
	=
	\left\{
	\begin{array}{c}
		m_b-\hm														\\
		m_bL_{CM}-\frac{1}{2}\hm(L_1-L_2)   \\
		\I_b-\frac{1}{3}\hm\left(L_2^2-L_1L_2+L_2^2\right)
	\end{array}
	\right\}
\end{eqnarray}
and measured parameters on the right hand side.
\Eref{eq:Matrix} written as~$\AM\mM=\BM$ has the least squares 
solution~$\mM=(\AM^T\AM)^{-1}\AM^T\BM$ and the distance~$D$ from this to 
a common solution is~$D=D(\tL_1,\tL_2,\I_b)=(\AM\mM-\BM)^T(\AM\mM-\BM)$.
Therefore, for any choice of two of the three parameters~$(\tL_1,\tL_2,\I_b)$ 
the third is determined by~$D=0$ and a common solution for~$\mM$
is then found by solving the normal equations~\eref{eq:Matrix}.
\Tref{tab:Solutions} shows such common solutions,~$\mM=\{\tm_1,\tm_2\}^T$,
\begin{table}[tbh]\footnotesize   
	\centering
		\begin{tabular}{cc||ccc|ccc|ccc}
			\multicolumn{2}{c||}{ }&
			\multicolumn{3}{c|}{$\I_b=1.70$kgm$^2$}&
			\multicolumn{3}{c|}{$\I_b=1.85$kgm$^2$}&
			\multicolumn{3}{c} {$\I_b=2.00$kgm$^2$}			\\
			$\tL_2$ 		& $L_2-\tL_2$ &
			$L_1-\tL_1$ & $\tm_1$ 	& $\tm_2$	&	
			$L_1-\tL_1$ & $\tm_1$ 	& $\tm_2$ & 
			$L_1-\tL_1$ & $\tm_1$ 	& $\tm_2$  						\\
			cm 					& cm				&
			cm 					& kg 				& kg 			&
			cm 					& kg 				& kg 			&
			cm 					& kg 				& kg  		     				\\\hline
			20 					& 5					&
			19.2				& 1.63 			& 0.02		& 
			9.8					& 1.48 			& 0.17		&
			0.4					& 1.37 			& 0.28		  					\\
			25 					& 0						&
			19.2				& 1.63 			& 0.02		& 
			10.3				& 1.50 			& 0.15		& 		
			1.4					& 1.39 			& 0.26   	
		\end{tabular}
	\caption{Model parameters that satisfy~\eref{eq:Matrix}.}
	\label{tab:Solutions}
\end{table}
for three values of~$\I_b$, that cover the experimental uncertainty 
in~\tref{tab:Param}, and two choices of~$\tL_2$.
When~$\I_b=1.85$kgm$^2$, 
most of the mass~$\tm_1+\tm_2=1.65$kg is placed at a reasonable distance 
of~$L_1-\tL_1\simeq10$cm from the long end of the beam irrespective 
of~$\tL_2$.
For the two other values of~$\I_b$, 
this distance is either too long~($\simeq20$cm) or too short~$(\simeq1$cm). 
This result consolidates the value and uncertainty of~$\I_b$ in~\tref{tab:Param}.
\section{Range}
\subsection{Range in vacuum}\label{app:VacRan}
The vacuum range~$R_v$ measured from the horizontal position of the fulcrum to 
impact on ground at the level of the trough is
\begin{eqnarray}\label{eq:R}
 R_v=\frac{v_i^2}{2g}
 \left(\sin(2A)+\sqrt{\sin^2(2A)+\frac{8gy_i}{v_i^2}\cos^2(A)}~\right)-x_i.
\end{eqnarray}
Here $\vM_i$ is the initial velocity at release and~$v_i=|\vM_i|$ the speed,
$(x_i,y_i)$ the horizontal position and height, and~$A$ the angle of~$\vM_i$ 
relative to the horizontal plane.
The parameters~$x_i$ and~$y_i$ are given by \eref{eq:rP} in terms of~$\ta$ 
and~$\phi$, and~$\vM_i$ and~$A$ follow by differentiation:
\begin{eqnarray}
		\vM_i&=
		(L_1\cos\ta~\dot\ta+L_4\sin\phi~\dot\phi)\eM_x+
		(L_1\sin\ta~\dot\ta-L_4\cos\phi~\dot\phi)\eM_y \nonumber\\
		A&=\arctan
		\left(-\frac{v_{iy}}{v_{ix}}\right)
		\quad
		\mathrm{with}
		\quad
		v_{ix}<0.
		\nonumber
\end{eqnarray}
\subsection{Range in air}\label{app:AirD}
The aerodynamic force is modeled by
\begin{eqnarray}\label{eq:AeroForce}
	\FM_a=-\frac{1}{2}C\rho_a A_Pv^2\eM_v.
\end{eqnarray}
Here~$C\simeq1/2$ is an empirical constant that depends on the shape and 
roughness of the projectile with aerodynamic cross section~$A_P$ and speed~$v$.
The air density is~$\rho_a$ and~$\eM_v$ is a unit vector along the path.

We assume that the aerodynamic loss is much smaller than the mechanical energy
of the projectile and that the vacuum range is much larger than the size of the engine. 
The loss~$Q$ can then be estimated by perturbation theory applied to a symmetric ballistic 
vacuum trajectory without losses that starts at~$(-R_v/2,0)$ and ends at~$(R_v/2,0)$
\begin{eqnarray}\nonumber
		Q=\int_{-T/2}^{T/2}(\FM_a\cdot v\eM_v)dt
					=C\rho_a A_P\int_0^{T/2}v^3dt,
\end{eqnarray}
where
\begin{eqnarray}\nonumber
		\frac{T}{2}=\frac{v_i\sin(A)}{g} 
		\quad\mathrm{and}\quad 
		v^2=(v_i\cos(A))^2+(gt)^2.
\end{eqnarray}
With the substitution~$t=(v_i\cos A/g)\tau$ one finds
\begin{eqnarray}\label{eq:Q}
		Q=C\rho_a A_P        
		\frac{v_i^4}{g}\cos^4(A)
		\int_0^{\tan(A)}    
		\left(1+\tau^2\right)^{3/2}d\tau, 
\end{eqnarray}
and the relative loss is
\begin{eqnarray}\label{eq:Q_E}
		\frac{Q}{E}&=4\pi\left(\frac{4\pi}{3}\right)^{-2/3}
		\frac{C\rho_a}{g\rho_s^{2/3}}~\frac{E}{m^{4/3}}~\frac{\F}{(\tan^2(A)+1)^2}										
		,
\end{eqnarray}
where~$\F$ is the definite integral in~\eref{eq:Q}.
We have also used~$E=(1/2)mv_i^2$,~$A_P=\pi r_p^2$ and~$m=\rho_s(4\pi/3)r_p^3$,
where~$\rho_s$ is the density of stone.
The function~$\F$ was evaluated analytically, and with~$x=\tan(A)$ the angular
dependence in~\eref{eq:Q_E} is
\begin{eqnarray}\nonumber
		\frac{\F(x)}{(x^2+1)^2}=
		\frac{3\sinh^{-1}(x)+x(2x^2+5)\sqrt{x^2+1}}{8(x^2+1)^2}.
\end{eqnarray}
This is almost constant in the interval from~$30^\circ$ to~$50^\circ$
and is replaced by the value at~$A=40^\circ$. 
With~$\rho_a=1.225$kg/m$^3$,~$\rho_s=2700$kg/m$^3$
and~$C=0.5$ we find in SI units
\begin{eqnarray}\label{eq:Q_E_fin}
		\frac{Q}{E}=
		6.2\cdot10^{-4}\frac{E}{m^{4/3}}=	
		3\cdot10^{-3}\frac{R}{m^{1/3}}.
\end{eqnarray}

The continuing loss of mechanical energy along the path shortens the range
such that approximately
\begin{eqnarray}\label{eq:DR_R}																			
		\frac{\D R}{R}=\frac{1}{2}\frac{Q}{E} 		
		,
\end{eqnarray}
which is the vacuum range calculated with the reduced initial 
velocity~$(v_i+v_f)/2$, where~$mv_f^2=2(E-Q)$.
The expressions~\eref{eq:Q_E_fin}	and~\eref{eq:DR_R} hold when~$Q$ is 
somewhat less than~$E$ and slightly overestimate the reductions because 
they are calculated with a trajectory that is too long and the velocities 
along the trajectory are also too large.


\section*{References}
\begin{thebibliography}{10}
%

\bibitem{ref:PVH} 
	Hansen, P. V., Acta Archaeologica, \textbf{63}, (1992), pp189-268

\bibitem{ref:Chev} 
	Chevedden, P. E.  (2000), 
	Dumbarton Oaks Papers, 54. Washington D.C.
	
\bibitem{ref:TS} 
	Saimre T., Estonian Journal of Archaeology, (2006), \textbf{10}, 1, pp61-80

\bibitem{ref:Chev_et_al} 
	Chevedden, P. E., Eigenbrod, L., Foley, V., and Soedel, W. (1995),
	Sci. Am. \textbf{273}, pp66-71.
	
\bibitem{ref:MSF1} 
	Fulton M. S., Chrissis N. G., Phillips J. and Kedar B. Z. 
	Crusades, (2017), \textbf{16}, pp33-53 
	
\bibitem{ref:MSF2} 
	Fulton M.~S., Artillery in the Era of the Crusades: 
	Siege Warfare and the Development of Trebuchet Technology.
	History of Warfare, vol. 122, BRILL (2018)
	
\bibitem{ref:Horsdal}
	Horsdal E., arXiv e-prints, arXiv:2303.01306

\bibitem{ref:O'Connell}	
	O'Connell J.,
	The Physics Teacher 39, 471 (2001); https://doi.org/10.1119/1.1424595
	
\bibitem{ref:Denny_2005}
	Denny M., Eur. J. Phys. 26 561	(2005)
	
\bibitem{ref:Denny_2009}
	Denny M., The Physics Teacher 47, 574 (2009); https://doi.org/10.1119/1.3264587

\bibitem{ref:Christo}
Christo Z., Phys. Educ. 52 013010 (2017)
	
\bibitem{ref:VTreb}	
	https://virtualtrebuchet.com/simulator
 
\end{thebibliography}
\end{document}